\documentclass[twoside,twocolumn,9pt]{article}
\usepackage{extsizes}
\usepackage[super,sort&compress,comma]{natbib} 
\usepackage[version=3]{mhchem}
\usepackage[left=1.5cm, right=1.5cm, top=1.785cm, bottom=2.0cm]{geometry}
\usepackage{balance}
\usepackage{mathptmx}
\usepackage{sectsty}
\usepackage{verbatim}
\usepackage{graphicx} 
\usepackage{lastpage}
\usepackage[format=plain,justification=justified,singlelinecheck=false,font={stretch=1.125,small,sf},labelfont=bf,labelsep=space]{caption}
\usepackage{float}
\usepackage{fancyhdr}
\usepackage{fnpos}
\usepackage[english]{babel}
\addto{\captionsenglish}{%
  
}
\usepackage[normalem]{ulem}
\usepackage{array}
\usepackage{droidsans}
\usepackage{charter}
\usepackage[T1]{fontenc}
\usepackage[usenames,dvipsnames]{xcolor}
\usepackage{setspace}
\usepackage[compact]{titlesec}
\usepackage{hyperref}

\usepackage{epstopdf}

\definecolor{cream}{RGB}{222,217,201}

\begin{document}

\pagestyle{fancy}
\thispagestyle{plain}
\fancypagestyle{plain}{
\renewcommand{\headrulewidth}{0pt}
}

\makeFNbottom
\makeatletter
\renewcommand\LARGE{\@setfontsize\LARGE{15pt}{17}}
\renewcommand\Large{\@setfontsize\Large{12pt}{14}}
\renewcommand\large{\@setfontsize\large{10pt}{12}}
\renewcommand\footnotesize{\@setfontsize\footnotesize{7pt}{10}}
\makeatother

\renewcommand{\thefootnote}{\fnsymbol{footnote}}
\renewcommand\footnoterule{\vspace*{1pt}%
\color{cream}\hrule width 3.5in height 0.4pt \color{black}\vspace*{5pt}} 
\setcounter{secnumdepth}{5}

\makeatletter 
\renewcommand\@biblabel[1]{#1}            
\renewcommand\@makefntext[1]%
{\noindent\makebox[0pt][r]{\@thefnmark\,}#1}
\makeatother 
\renewcommand{\figurename}{\small{Fig.}~}
\sectionfont{\sffamily\Large}
\subsectionfont{\normalsize}
\subsubsectionfont{\bf}
\setstretch{1.125} 
\setlength{\skip\footins}{0.8cm}
\setlength{\footnotesep}{0.25cm}
\setlength{\jot}{10pt}
\titlespacing*{\section}{0pt}{4pt}{4pt}
\titlespacing*{\subsection}{0pt}{15pt}{1pt}

\makeatletter 
\newlength{\figrulesep} 
\setlength{\figrulesep}{0.5\textfloatsep} 

\newcommand{\topfigrule}{\vspace*{-1pt}%
\noindent{\color{cream}\rule[-\figrulesep]{\columnwidth}{1.5pt}} }

\newcommand{\botfigrule}{\vspace*{-2pt}%
\noindent{\color{cream}\rule[\figrulesep]{\columnwidth}{1.5pt}} }

\newcommand{\dblfigrule}{\vspace*{-1pt}%
\noindent{\color{cream}\rule[-\figrulesep]{\textwidth}{1.5pt}} }

\makeatother

\twocolumn[
  \begin{@twocolumnfalse}
    {
    }\par
\vspace{1em}
\sffamily
\begin{tabular}{m{4.5cm} p{13.5cm} }

   & \noindent\LARGE{
    \textbf{Mixing properties of bi-disperse ellipsoid assemblies: Mean-field behaviour in a granular matter experiment$^\dag$ \linebreak {\small (Final version published in Soft Matter, Issue 5, 2023, see \href{https://doi.org/10.1039/D2SM00922F}{https://doi.org/10.1039/D2SM00922F})}
    }
  } \\
\vspace{0.3cm} & \vspace{0.3cm} \\

 & \noindent\large{F.M.\ Schaller,\textit{$^{a,b}$} H.\ Punzmann,\textit{$^{c}$}, G.E.\ Schr\"oder-Turk,\textit{$^{\ast,d,c,a}$} and M.\ Saadatfar,\textit{$^{\ast,c,e}$}} \\

& \noindent\normalsize{
The structure and spatial statistical properties of amorphous ellipsoid assemblies have profound scientific and industrial significance in many systems, from cell assays to granular materials. 
This paper uses a fundamental theoretical relationship for mixture distributions to explain the observations of an extensive X-ray computed tomography study of granular ellipsoidal packings.
We study  { a size-}bi-disperse mixture of { two types of} ellipsoids of revolutions that have the { same aspect ratio of $\alpha\approx 0.57$ and differ in size, by about 10\% in linear dimension}, and { compare these to mono-disperse systems of ellipsoids with the same aspect ratio}. Jammed configurations with a range of packing densities are achieved by employing different tapping protocols. We numerically interrogate the final packing configurations by analyses of the local packing fraction distributions calculated from the Voronoi diagrams.
Our main finding is that the bi-disperse ellipsoidal packings studied here can be interpreted as a mixture of two uncorrelated mono-disperse packings, insensitive to the compaction protocol. 
Our results are consolidated by showing that the local packing fraction shows no correlation beyond their first shell of neighbours in the binary mixtures. We propose a model of uncorrelated binary mixture distribution that describes the observed experimental data with high accuracy.
This analysis framework will enable future studies to test whether the observed mean-field behaviour is specific to the particular granular system 
{
or the specific parameter values 
}
studied here or if it is observed more broadly in other bi-disperse non-spherical particle systems. 

} \\

\end{tabular}

 \end{@twocolumnfalse} \vspace{0.6cm}

  ]

\renewcommand*\rmdefault{bch}\normalfont\upshape
\rmfamily
\section*{}
\vspace{-1cm}


\footnotetext{\textit{$^{a}$~Friedrich-Alexander Universit\"at Erlangen-N\"urnberg, Institut f\"ur Theoretische Physik, Staudtstr.\ 7B, 91058 Erlangen, Germany}}
\footnotetext{\textit{$^{b}$~Karlsruhe Institute of Technology (KIT), Institut f\"ur Stochastik, 76131 Karlsruhe, Germany}}
\footnotetext{\textit{$^{c}$~The Australian National University, Research School of Physics, Canberra ACT 2601, Australia}}
\footnotetext{\textit{$^{d}$~Murdoch University, College of Science, Technology, Engineering and Mathematics, 90 South St, Murdoch WA 6150, Australia}}
\footnotetext{\textit{$^{e}$~The University of Sydney, School of Civil Engineering, NSW 2006, Australia}}
\footnotetext{\textit{$^{\ast}$~Authors for correspondence: Gerd Schr\"oder-Turk (g.schroeder-turk@murdoch.edu.au) and Mohammad Saadatfar (mohammad.saadatfar@sydney.edu.au )}}

\footnotetext{\dag~Electronic Supplementary Information (ESI) available: [details of any supplementary information available should be included here]. See DOI: 10.1039/cXsm00000x/}



Packing properties are integral for understanding the behaviour of many-particle assemblies, including granular matter, glasses, liquids, gels, cell tissues, self-assembled micellar systems, etc. Among this variety of systems, athermal granular materials of hard particles are conceptually particularly simple systems that nevertheless show a range of complex emergent phenomena. As such (and aside from their scientific and industrial relevance in their own right), they lend themselves as toy models for the study of specific aspects of packing phenomena -- as, for example, the mixing properties addressed here. 

Realistic granular materials are complex mixtures of particles, with 
variations in particle shape, particle volume or composition of the 
different constituents. The construction of conceptual models for such complex systems is an important step in the scientific understanding of their properties. 

The simplest model are mono-disperse (equal 
size) sphere packings
\cite{bernal1960packing,bernal1964bakerian,silbert2002geometry,Aste2005}. 
This simplest model can be extended in various ways, such as by introducing size polydispersity or changes to the particle shape. 

Spherical systems with particle size distribution (polydispersity or bi-dispersity) have been extensively studied, specifically in the context of glass forming systems \cite{charbonneau2014p,seguin2016experimental,parisi2010mean}, as well as in Apollonian packings \cite{PRLGary} \cite{Anishchik1995Apollonian3D}\cite{Reis2012}. Numerous studies have addressed random close packing limits and jamming phenoma in bi- or polydisperse spheres \cite{BaranauPolydisperse,PhysRevE.60.7098,PhysRevE.100.042906,PhysRevE.80.051305,doi:10.1073/pnas.2021794118,LiWeiwei,PhysRevResearch.3.L032042,doi:10.1063/1.1511510,PhysRevE.82.011403,10.21468/SciPostPhys.3.4.027}. 

{ Monodisperse} granular (and related) systems with non-spherical shapes have been extensively studied, too, with particle shapes ranging from convex ellipsoids\cite{DelaneyHiltonCleary2011}, polyhedra and super-cubes to concave particles\cite{PhysRevLett.119.028003}. Ellipsoidal packings and configurations have received attention as models for granular systems \cite{Donev2004,RevModPhys.90.015006,Chen2021,PhysRevLett.102.255501,Zeravcic_2009,Delaney_2010,C3SM52047A}. Ellipsoidally shaped particles, either as configurations or particle shapes, have a wide ranging relevance for other systems such as \cite{Rollere2021,Li2018,Deng2019,Han2006,Chen2018,Jeffery1922,Burke2015,Davies2014,Lovric2019,PhysRevLett.92.255506,BezrukovStoyan2006,Stoyan2021}.  
Structural properties of non-spherical packings have been addressed theoretically \cite{baule2018edwards,torquato2010jammed,jiao2011maximally,baule2013mean} and by experiments \cite{man2005experiments,neudecker2013jammed,schaller2015local,jaoshvili2010experiments}. 

Polydisperse aspherical particle packings combine non-spherical particle shapes with variations in particle size or shape. The bi-disperse ellipsoid packings studied here, where both particle types have the same ellipsoidal shape but different size, are an example of this type. 
{
Yuan {\it et al.} have conducted an extensive computational study of polydisperse particle packings and have found a universal behaviour of the average Voronoi cell volume (rather than of the cell volume distributions) as function of an (effective) particle size \cite{yuan2020} (see also earlier work \cite{YuanPRE2018}).
}
{
The deformation behaviour of bi-disperse ellipsoid assemblies has been studied \cite{NG20045871,NG2009748}, as well as the influence of particle shape and angularity on the properties of granular materials.
}

\begin{figure}
  \includegraphics[height=3.7cm]{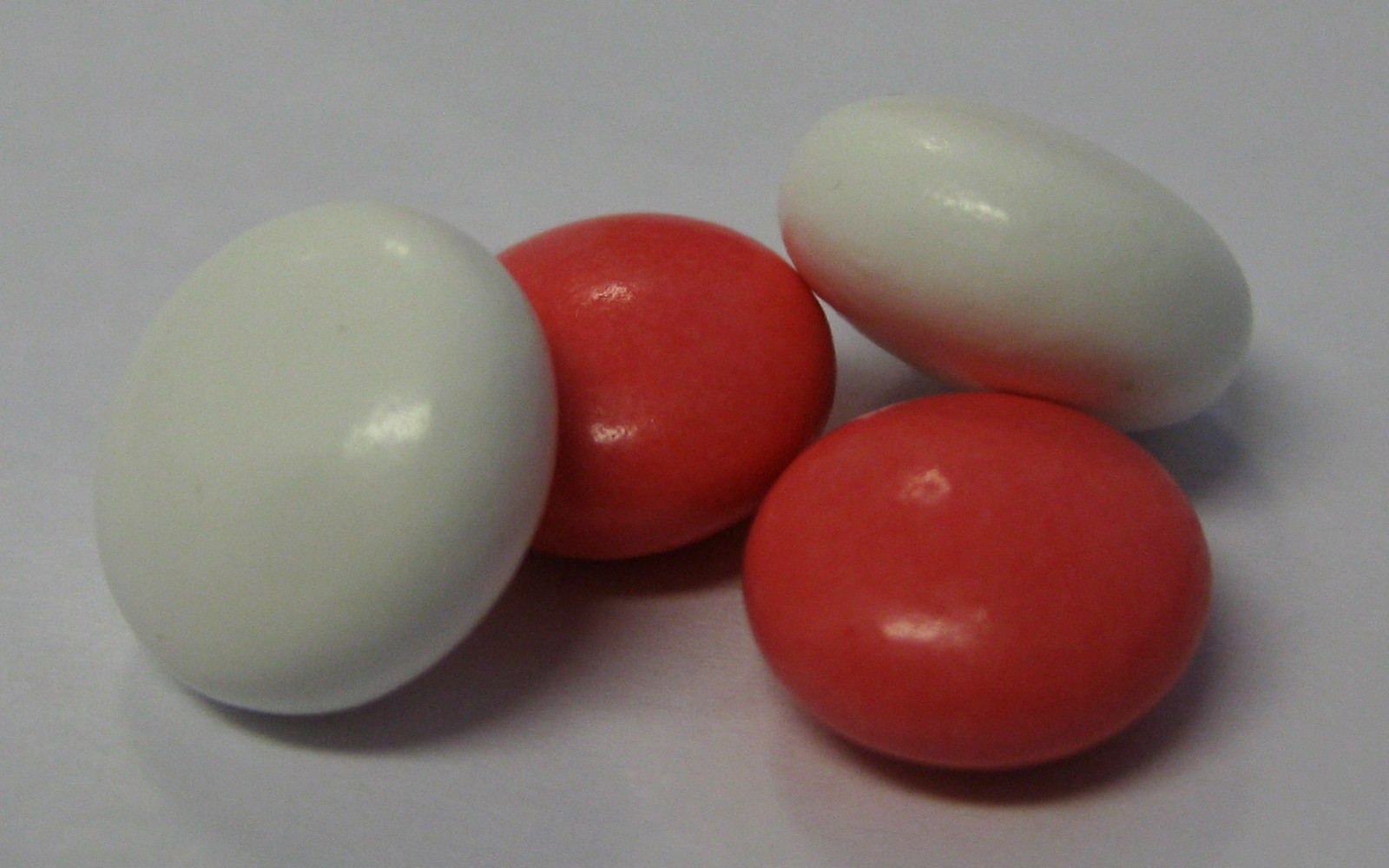}
  \includegraphics[height=3.7cm]{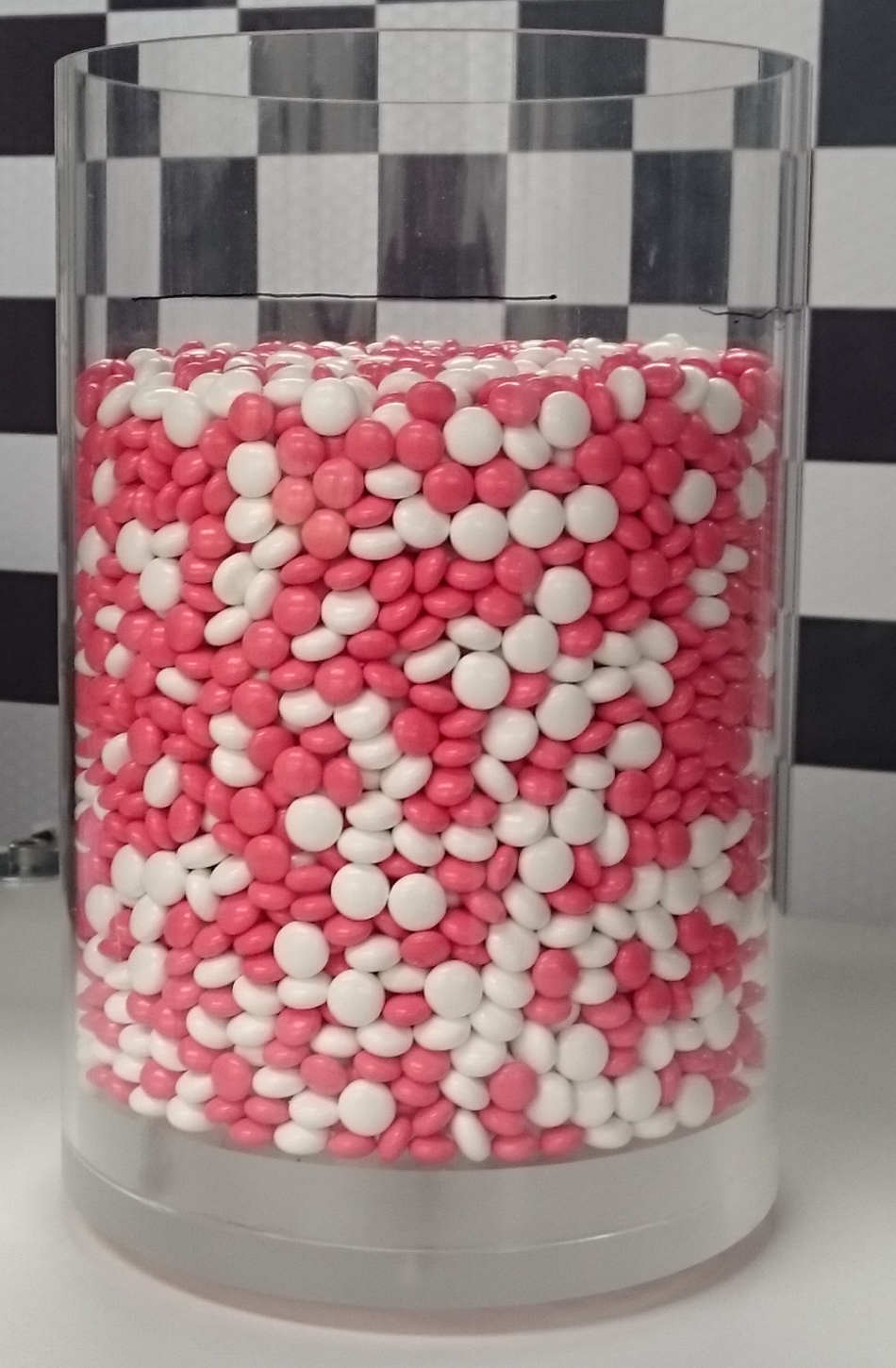}\\[0.05cm]
    \includegraphics[height=4.77cm]{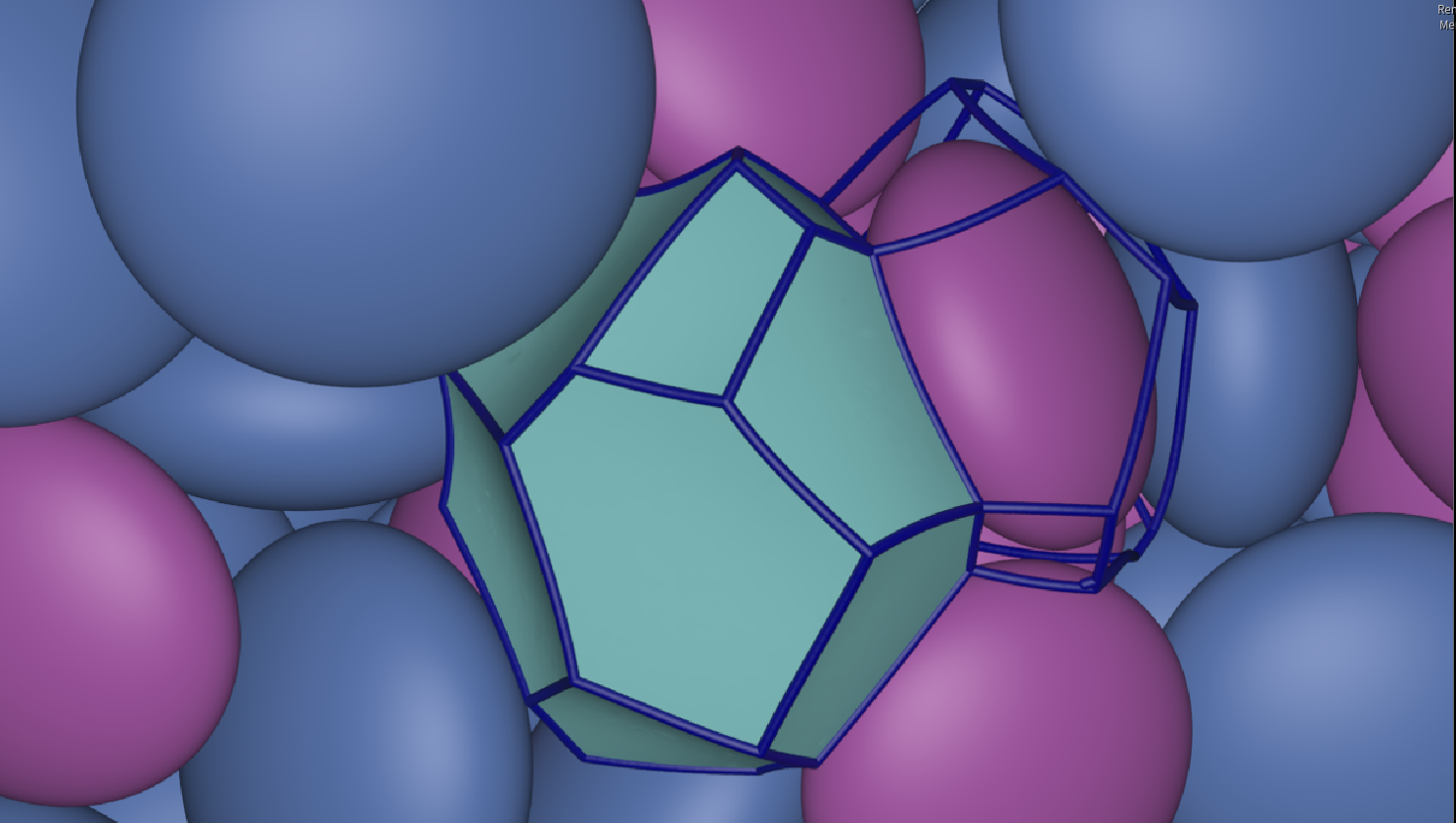}
  \caption{The top row shows photographs of the two types of pharmaceutical placebo pills with two different aspect ratios used in this study and of a bi-disperse packing of ellipsoids created from a mixture of such particles. The bottom image shows a computer-generated illustration of a small section of a larger bi-disperse ellipsoid packing, together with the Voronoi cell of two specific particles. The illustration is generated from position data of ellipsoid particles extracted from a tomography image. Voronoi cells are obtained by the Set-Voronoi algorithm \cite{Schaller2013PhilMag,pomelo} which, as is visible in the picture, can lead to curved faces and edges.}
\label{fig:photos-visualisation}
\end{figure}

Due to the large number of parameters such as particle geometry, particle connectivity, particle-particle law of interaction and particle size distribution, the characterisation of particle packings is very complex. The combined effect of many particle scale details (i.e particle shape, size, surface topography, friction, particle connectivity etc.) is contained inside the free volume surrounding each particle known as the Voronoi cell (polygon) \cite{Schaller2013PhilMag,Aste2005}.
The fraction of space occupied by each particle inside its Voronoi cell is the local packing fraction
\begin{equation}
  \Phi_l = \tfrac{V_\textnormal{particle}}{V_\textnormal{cell}}.
\end{equation}
Statistical analysis of $\Phi_l$ and geometrical characterisation of the shape of Voronoi polygons in static granular assemblies have revealed important universalities and predictive behaviours of granular media, unaffected by packing protocols \cite{Aste2005,Aste2006,SchroederTurk2010EPL,Schaller2015EPL}.

This paper presents the first experimental observation of ellipsoidal packings (mono- and bi-disperse) where the effect of particle shape and size on structural correlation is investigated.
We present experimental results obtained by x-ray tomography investigating the structure of mono- and bi-disperse ellipsoidal packings. We produce a range of packing densities via different tapping protocols, where each configuration between successive taps is a jammed state. 

Our approach gives detailed insights into the 3D structure of assemblies of ellipsoids, which is here used to understand the statistical mixing properties (those related to the presence of two particle sizes) in detail.

\section{Experiment \& analysis}

In our experiments, we use oblate ellipsoids (ellipsoids of revolution with two equal axes) with aspect ratio $\alpha \approx 0.57$ and half-axis $a:a:\alpha a$.
Our particles are pharmaceutical placebo pills of two sizes, small ones with $a = 4.45$ mm and big ones with $a = 5.10$ mm. 
The smaller particles are of the same particle type and size as used in \cite{Schaller2015EPL},  the larger particles are particles of the same density and same smooth (low-friction) surface coating.
Aside from their size, the physical properties of the big and small particles are the same in terms of density and surface properties. 
The variations among particles of the same type are very small, indicating well-defined particle types; the standard deviation of the linear sizes is $<2\%$ of the particle size, for each of the two types.

\begin{figure}
  \includegraphics[width=0.48\textwidth]{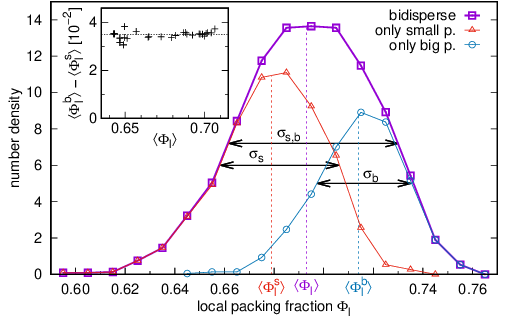}
  \caption{Local packing fraction distribution of a bi-disperse packing ($\langle\Phi_l\rangle = 0.693$)
           and the mono-disperse components when only small/big particles are considered ($\langle\Phi_l^s\rangle = 0.679$, $\langle\Phi_l^b\rangle = 0.714$).
  Inset: Difference of the mean local packing fraction of the big and small particles in all our bi-disperse packings is approximately constant
  ($\Delta \Phi = \langle\Phi_l^b\rangle - \langle\Phi_l^s\rangle \approx 0.035$)
  .}
  \label{fig:local_pf_distr}
\end{figure}

The small particles are red, whereas the big particles are white (but are depicted as blue in Fig.\ \ref{fig:photos-visualisation}(bottom) and in the inset of Fig.\ \ref{fig:pf_sigma}. In Figs.\ \ref{fig:local_pf_distr} and \ref{fig:pf_sigma}, we use red symbols to refer to properties of the small particles and blue symbols to refer to those of the big particles.

For this study, we have prepared, by the method described below, and individually imaged by tomography 26 packings of mono-disperse assemblies, where all particles are either the big particles or the smaller particles (These are complemented by a further earlier data sets from \cite{Schaller2015EPL}); each orange data point in Fig.\ \ref{fig:pf_sigma} represents one such packing. We have also prepared 26 packings of bi-disperse mixtures of the smaller particles and the larger particles; each of the purple data points in Fig.\ \ref{fig:pf_sigma} represent a packing.

For the bi-disperse packings, we use a particle mixing ratio of 2:3 (that is, a fraction of $m=2/5$ of the particles are big, and $(1-m)=3/5$ are small). This results in the same total volume of each particle type in the container ($\sum_{N_b} V_\textnormal{particle}^b = \sum_{N_s} V_\textnormal{particle}^s$), where $N_b$ and  $N_s$ are the number of big and small particles.

Initially, a loose packing is prepared by pouring the particles into a cylindrical container on a rotational stage and a subsequent slow removal of an assembly of horizontal grids, which previously have been placed inside the container.
After this initial loose packing is prepared, we create packings of different packing fractions by vertical tapping using different tapping protocols.
Each configuration between successive taps is a jammed configuration and the system is given time to relax under gravity before the application of the next tap.

The final packings are imaged by X-ray tomography. 
{
Tomographic imaging was carried out using a cone-beam x-ray micro-CT instrument that was designed and built in-house at the Australian National University. A flat panel detector (pixel size 0.14 $\mu m$) and micro-focus x-ray source (acceleration voltage 100 kV and current 100 $\mu$A) supplied by Hamamatsu, were used for tomographic imaging of packings. The samples were placed on a motor controlled rotating stage and radioscopic projections were taken after each degree of rotation. In total, around 10,000 projections were obtained for a complete rotation of the sample. We employ an optimally prescribed set of viewing angles that follow a space-filling helical trajectory. The exposure time was 1 s and two images per projection were taken to reduce the noise. The total imaging time was approximately 6 h for each sample. The x-ray source has a cone beam geometry which allowed magnification; as a result, we achieved voxel resolution of 56 $\mu m$.
}

The particle positions and orientations are detected in the reconstructed image by a method based on a watershed algorithm \cite{Schaller2013AIP}.

The local packing fraction $\Phi_l$ of the particles is extracted by calculating the Set-Voronoi cells of the particles, { based on a Voronoi algorithm of a mesh discretisation of the particle surfaces} \cite{Schaller2013PhilMag,pomelo}. 

Details of the preparation setup, tomographic imaging and particle detection can be found in Ref. \cite{Schaller2017bidisperse}.

{
We have verified that our packings are, both upon preparation and after tapping, of sufficient mixing homgeneity in both radial and vertical direction. We have analysed the vertical and radial mixing ratio profiles; that is, the mixing ratios $m(h)$ as the ratio of the number of large particles to the number of all particles in a layer of small height $\Delta h$ as a function of position $h$ along the vertical direction and $m(r)$ as the ratio of large particle numbers to all particles in a cylindrical layer of small width $\Delta r$ as a function of the radial position $r$. We see a small systematic variation of $\langle m(r)\rangle$ as a function of $r$ (when averaged over all packings); near the middle of the sample $\langle m(r)\rangle \approx 0.42$ whereas near the cylinder surface $\langle m(r)\rangle \approx 0.38$. Along the vertical direction, we see a small systematic variation of $\langle m(h)\rangle$ where it increases from roughly 0.37 at the base of the cylindrical container to 0.43 at the upper end of the container. These effects are small effect and are effects of the initial pouring; we see no greater systematic variations in $m(r)$ or $m(h)$ after tapping than we do after the initial preparation. The statistical variations between different samples are slightly larger than the systematic variations described above.
We have not expressly verified the absence of orientational ordering in the samples; however, we do not expect any long-range ordering based on the fact that the mono-disperse system of the same particles do not show any significant ordering \cite{schaller2015local}, aside from a very weak preference for alignment with the direction of gravity. Given the low aspect ratios, one would not expect any ``smectic'' ordering). 
}

We describe each distribution by the mean local packing fraction $\langle \Phi_l \rangle$ and the width of the distribution $\sigma$.
In mono-disperse packings, the global packing fraction $\Phi_g$, which is the fraction of space occupied by the particles in the whole packing, is the harmonic mean of the local packing fractions $\Phi_g=\frac{N}{\sum 1/\Phi_l}$.

Fig.~\ref{fig:local_pf_distr} shows the local packing fraction distribution for a binary packing (open squares).

In order to compare mono- and bi-disprese packings, we use the computed mean local packing fraction $\langle \Phi_l \rangle = 1/N \sum_N \Phi_l$, because the commonly used quantity of the global packing fraction, $\langle \Phi_g\rangle$, cannot be extracted solely from the local packing fractions $ \Phi_l$ in bidisperse packings. This is due to the fact that to calculate $\Phi_g$ in bi-disperse packings, $V_\textnormal{particle}$ and $V_\textnormal{cell}$ for each particle are needed.

Hence, in the following, we use the mean local packing fraction $\langle \Phi_l \rangle = 1/N \sum_N \Phi_l$.
Furthermore, for bi-disperse packings, we define the mean local packing fraction of the big particles $\langle \Phi_l^b \rangle$ by considering only the local packing fractions $\Phi_l^b$ of the big particles:
\begin{equation}
  \langle \Phi_l^b \rangle =
  \frac 1 {N_b} \sum_{N_b} \Phi_l^b = 
  \frac 1 {N_b} \sum_{N_b} \frac{V_\textnormal{particle}^b}{V_\textnormal{cell}^b}
\end{equation}
with $N_b$ being the number of big particles.
$\langle \Phi_l^s \rangle$ for the small particles is defined accordingly.
The two components of a bi-disperse distribution are shown in Fig.~\ref{fig:local_pf_distr}.

\section{Results}

\begin{figure}
  \begin{picture}(244,212)    
  \put(0,0){\includegraphics[width=0.48\textwidth]{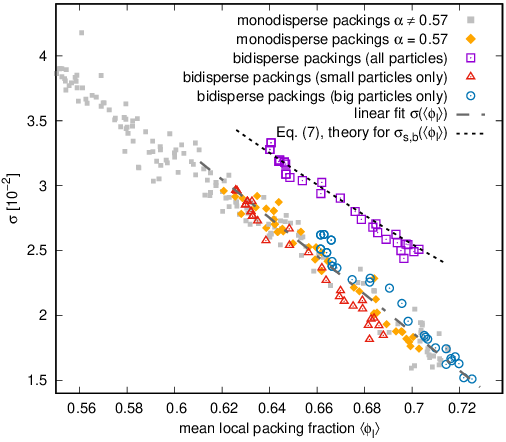}}
  \put(35,30){\includegraphics[width=0.16\textwidth]{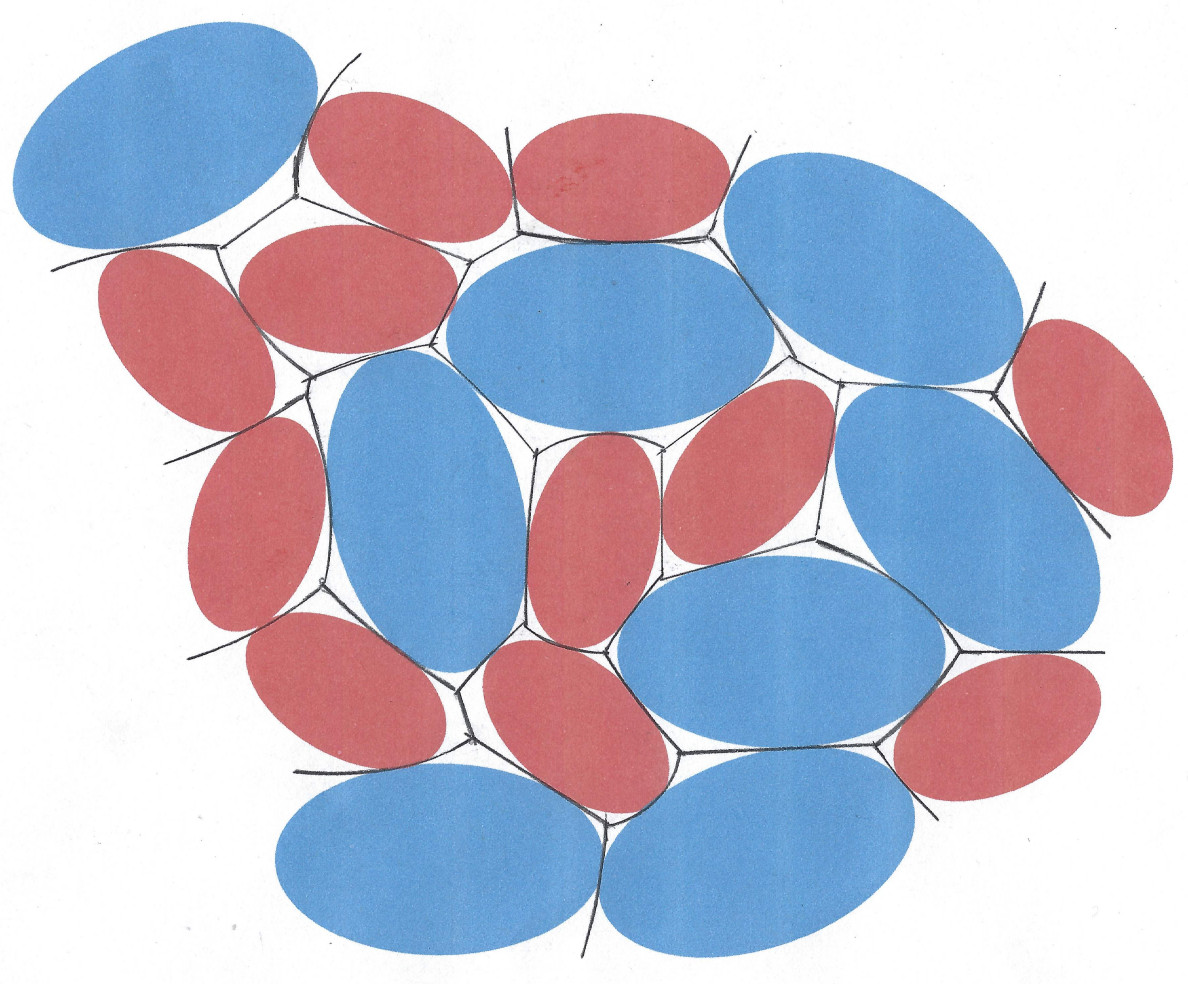}}
  \end{picture}
  \caption{Width $\sigma$ of the local packing fraction distributions of monodisperse and bidisperse packings.
           The open triangle and circle points are the widths of the distributions of bidisperse packings if only small or big particles are considered.
           The width of these distributions matches the ones of monodisperse packings. In addition, packings from Ref \cite{Schaller2015EPL} are included (solid squares). The illustration of particles merely facilitates color identification in the plots and is not to scale.}
  \label{fig:pf_sigma}
\end{figure}

The main result of this work is presented in Fig.~\ref{fig:pf_sigma}, which shows the widths of the local packing fraction distributions $\sigma$ plotted as a function of the mean local packing fraction $\langle\Phi_l\rangle$. This is the central observation to demonstrate that the packing fraction distribution of our bidisperse packings, can be understood, to first order, as the mixture distribution of two uncorrelated monodisperse packings.)

In Fig.~\ref{fig:pf_sigma} we see that the mono-disperse packings of ellipsoids with $\alpha = 0.57$ used in this work (orange diamonds) are in good agreement with mono-disperse ellipsoid packings from simulations and experiments of other aspect ratios from Ref. \cite{Schaller2015EPL} (solid squares).
Bi-disperse packings (open squares) show larger values for  $\sigma_{s,b}$ i.e.~they have a wider distribution than mono-disperse packings with the same mean packing fraction.
Surprisingly, the widths of the mono-disperse components of the bi-disperse packings
$\sigma_s = \sigma[\Phi_l^s]$ and $\sigma_b = \sigma[\Phi_l^b]$
(open triangle and circle) as a function of the mean local packing fraction follow the same curve as the data for mono-disperse packings (solid squares and diamonds).

All data points in the relevant range, [0.61,0.73], collapse onto a linear fit function (dash-dotted fit in Fig.~\ref{fig:pf_sigma}):

\begin{equation}
  \sigma\left(\langle\Phi_l\rangle\right) = 0.122 - 0.147 \cdot \langle\Phi_l\rangle
  \label{eq:sigmapf}
\end{equation}
We now assume that the local packing fraction distribution of a binary packing $\sigma_{s,b}$ is the mixture distribution of two uncorrelated mono-disperse packings. The second moment of the mixture distribution with two components is given by
\begin{equation}
  \sigma_{s,b}^2 \left(\langle\Phi_l^b\rangle,\langle\Phi_l^s\rangle\right)
  = \sum_{i = b,s} w_i \Big( \sigma^2\left(\langle\Phi_l^i\rangle\right) + \langle\Phi_l^i\rangle^2 \Big) - \langle\Phi_l\rangle^2
  \label{eq:sigmatilde}
\end{equation}
with weights $w_i$ for the distributions of the big and small particles and
$\langle\Phi_l\rangle = w_b \langle\Phi_l^b\rangle + w_s \langle\Phi_l^s\rangle$, see \cite{MixtureDistribution}.
In our experiments, we use a mixing ratio of 2:3, so $w_b = 0.4$ and $w_s = 0.6$. 

By analyzing our packings, we find that 
%
$\Delta \Phi = \langle\Phi_l^b\rangle - \langle\Phi_l^s\rangle \approx 0.035$ is approximately constant,
see inset of Fig.~\ref{fig:local_pf_distr}.
This results in:
\begin{eqnarray}
  \langle\Phi_l^b\rangle & \approx \langle\Phi_l\rangle + w_s \tfrac{\Delta\Phi}{2} \nonumber\\
  \langle\Phi_l^s\rangle & \approx \langle\Phi_l\rangle - w_b \tfrac{\Delta\Phi}{2}
\end{eqnarray}
which leads to:
\begin{eqnarray}
  \sigma_{s,b}^2 \left(\langle\Phi_l\rangle\right)
  &=& w_s \Big(g\left(\langle\Phi_l\rangle - w_b \tfrac{\Delta\Phi}{2}\right)\Big) \nonumber\\
  &+& w_b \Big(g\left(\langle\Phi_l\rangle + w_s \tfrac{\Delta\Phi}{2}\right)\Big)
  - \langle\Phi_l\rangle^2
\end{eqnarray}
with $g(x) = \sigma^2(x) +x^2$.
By using Eq.~\ref{eq:sigmapf}, we get:
\begin{equation}
  \sigma_{s,b}^2\left(\langle\Phi_l\rangle\right) \approx
  0.0216 \langle\Phi_l\rangle^2 - 0.0359 \langle\Phi_l\rangle + 0.0152
\end{equation}
This is in very good agreement with our results, see dashed black line in Fig.~\ref{fig:pf_sigma}.
For our particles and mixing ratio, we find that the simple ansatz by considering the local packing fraction distribution of a binary packing as the mixture distribution of two mono-disperse ones is in very good agreement with the experimental results.
Reasons for the small deviations can be possible segregation effects during the tapping process or second order effects in the distribution, since we only consider $\langle\Phi_l\rangle$ and $\sigma$.


\subsection*{Local volume density correlations}

\begin{figure}
  \includegraphics[width=0.48\textwidth]{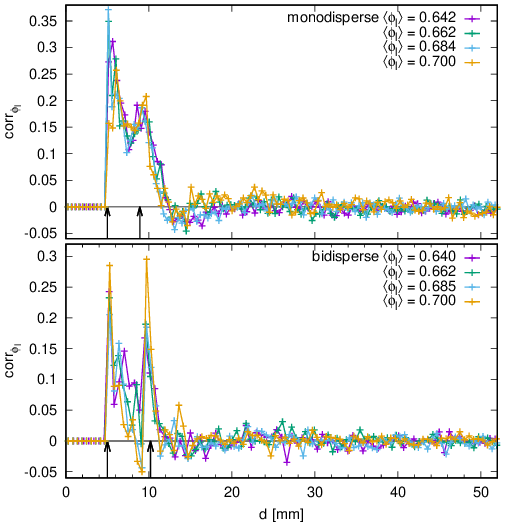}
  \caption{Local packing fraction correlation.
           The two arrows mark the minimal and maximal touching distance of two ellipsoids in the packing.}
  \label{fig:pf_corr}
\end{figure}

The good match of the model to the data shown in Fig.~\ref{fig:pf_sigma} suggests the simple uncorrelated mixture distribution model presented in Eq.~\ref{eq:sigmatilde} recovers our experimental results with reasonable accuracy. This may suggest the absence of long-range correlations in the local volume fractions. We investigate this by calculating the correlation between the local packing densities of all particles (big and small) using the two-point correlation function defined as: 

\begin{equation}
  \textnormal{corr}_ {\Phi_l} (d) = \frac{\left\langle\left(\Phi_l\left(r_1\right) - \langle \Phi_l \rangle\right) \cdot \left(\Phi_l\left(r_2\right) - \langle \Phi_l \rangle\right)\right\rangle }{\left(\sigma[\Phi_l]\right)^2}
\end{equation}
where  $d = r_1 - r_2$ is the distance between two arbitrary particles at position $r_1$ and $r_2$.

The packing fraction correlations in selected mono- and bi-disperse packings are shown in Fig.~\ref{fig:pf_corr}. The minimum and maximum touching distance between two ellipsoids are shown with arrows. Maximum correlations are measured around the minimum and maximum distance of touching ellipsoids with a drop in the first void between nearest ellipsoids.

In both mono- and bi-disperse packings, no long range correlation of the local packing fraction is observed. Beyond the first shell of neighbours, the correlation fluctuates around zero suggesting the presence of immediate neighbour correlation only. In dense packings we observe negative correlation in the short range regime consistent with previously reported results in dense bi-disperse 2D disk packings \cite{Zhao2012Disc}. 

\section{Discussion and Conclusion}
Although binary mixtures represent the simplest case of poly-disperse granular systems, they nonetheless exhibit a very important behavior of granular mixtures as reported in this work.


We showed that in bi-disperse ellipsoidal packings with moderate size differences, the mono-disperse components of the local packing fraction distribution characterized by the width of the Voronoi volume distribution, $\sigma$, are the same as in mono-disperse ellipsoidal packings. In other words, the  width $\sigma$ of the local packing fraction distribution of mono-disperse packings, and the width $\sigma$ obtained by separating the packing fractions associated with each particle size within the bi-disperse packing are indistinguishable. This implies that bi-disperse packings have the same volume distribution as the  mixture of two uncorrelated mono-disperse packings) and they follow the same trend as other mono-disperse packings, see Fig.~\ref{fig:pf_sigma}. 

We showed that a simple ansatz, that considers bi-disperse packings as a mixture of two mono-disperse packings (with no assumed correlations), is in a good  agreement with our experimental results.  

The master curve shown in Fig.~\ref{fig:pf_sigma} represents packing fractions in bi-disperse ellipsoidal packings that range from very loose ($\phi=0.623$) to dense ($\phi=0.724$). Therefore, the presented results are insensitive to the specifics of the tapping protocol. 

This confirms that  bi-disperse ellipsoidal packings can be considered as uncorrelated mixtures of two mono-disperse ellipsoidal packings with no spatial correlation between their local volume fractions.

The binary ellipsoidal packings considered in this study represent a larger class of physical systems where lack of long range correlation produces homogeneous and uncorrelated structures allowing the system's statistical properties (i.e. local volume distribution) to be described by the gamma distribution. Jammed packings of mono-sized spheres \cite{Aste2005, Aste2006, francois2013geometrical} and mixing of immiscible fluids via random stirring are examples of such systems \cite{duplat2008mixing}.

Recent studies of the packing structure of mono-disperse spheres have revealed that the structure is completely random (uncorrelated) and the jamming transition can be captured by the Erd\"os-R\'enyi network with no correlations \cite{morone2019jamming}. Furthermore, the mechanical backbone of such systems also shares similar trait with those of random networks \cite{saadatfar2017pore,PhysRevE.91.062202,PhysRevLett.113.148001,D1SM00774B}. This has led to the development of a mean-field theory for random close packings \cite{baule2018edwards} allowing a full description of system's configurational phase space. In the current study we provide experimental evidence showing that the packing structure of binary mixtures of lens-shaped ellipsoidal particles also follows this trend and thus it is valid to describe such systems by the mean-field theory of axi-symmetric particles \cite{baule2013mean}  and the random network theories for granular media with no correlations \cite{morone2019jamming}.

{
Our study has focused on distributions of local packing fractions (equivalent to free volumes), rather than how the average of those distributions depends on system parameters. The appendix provides a brief discussion of averages of our data in the context of the analysis by Yuan {\it et al} \cite{yuan2020}.  
}

{
We recognise that the key weakness of our analysis is the limitation to a single set of parameters for the mixing ratio of large and small particles ($2:3$), for the size ratio of the two particles of large to small particles ($5.10:4.45$) and for the particle aspect ratio ($\alpha=0.57$ for both particles). Our focus in this study was to establish, with statistical significance and confidence, the validity of the key result for this fixed set of these parameters while varying the average local packing fractions $\langle \phi_l\rangle$. Even our single-parameter study required a large experimental study comprising $>50$ tomographic analyses.

Future studies need to address the validity of the hypothesis, particularly of the master curve in Fig.~\ref{fig:pf_sigma}, for different values of the fixed parameters. It is clear that there is at least one limit where our hypothesis trivially works, namely the limit of mono-disperse particles, and it is likely that it also works for spherical particles with intermediate mixing ratios. However, it is also clear that there are regions or limits of the parameter space where the packings will cease to be homogeneous mixtures, notably the limit of large size difference between the particles. For such segregated systems, our analysis is not expected to be insightful. The focus of future research should establish, or disprove, the validity of our result in those parameter regions where the systems remain homogeneous, and address generalisations to non-ellipsoidal particle shapes and to particles with greater friction values.}

{
The analysis tools described in this article can be applied to other particle assemblies, both in terms of the numerical Voronoi analysis of configurations of polydisperse aspherical particles and in terms of the analytic mixing formula. Obvious examples would be the rich equilibrium phase diagram of aspherical liquid-crystalline systems, of fluids of aspherical particles, or of glassy systems such as in colloidal systems.   
}

{
\section*{Appendix: Analysis of averaged scaled free volumes as function of rescaled particle size}

\begin{figure}
  \includegraphics[width=0.48\textwidth]{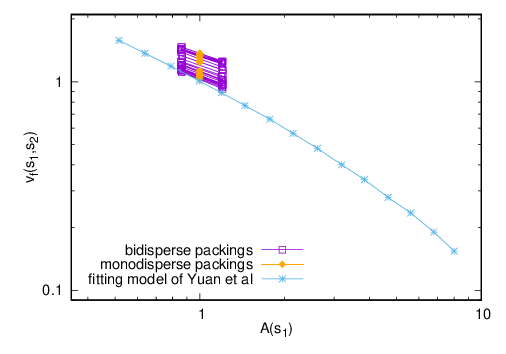}
  \caption{Comparison of data from our experiments for normalised free volumes $v_f$ and $A$ with the data presented by Yuan {\it et al} \cite{yuan2020}.}
  \label{fig:Yuan-averages}
\end{figure}

Our work has focused on distributions of Voronoi cell volumes, rather than on averages. Given the nature of the data presented here, it appears however useful to compare the data to the averages of free volume cells described by Yuan {\it et al} \cite{yuan2020}. We therefore compare averages (per particle type) of the particle sizes and Voronoi volumes to the relationships between normalised particle sizes and free volumes, in analogy to Figure 5 of the paper by Yuan {\it et al} \cite{yuan2020}. 

Their data, for size-polydisperse mixtures of particles of identical shape but vastly different sizes, covers a much broader range of polydispersities but includes particles of approximately the same shape as ours. Their data is for frictionless particles, and hence the random close packing limit, in contrast to our data. For each of the two particle types in our bidisperse packings and, separately, for our monodisperse packings, we analyse
\begin{equation}
v_f=\left(V_{c}/V_{p}-1\right)/\left(e_J-1\right)
\end{equation} 
as a function of 
\begin{equation}
A=D_v^2/\langle D_v^2\rangle.
\end{equation}
The inverse of the random close packing limit $e_J=1/\Phi_J$ is chosen as $\phi_j=0.71$ (as per data from Delaney {\it et al} for particles of aspect ratio $\alpha=0.57$ \cite{Delaney_2010}). For a particle of volume $V_{p(article)}$, $D_v=\left(3 V_p/(4\pi)\right)^{1/3}$ is the radius of the sphere with equal volume, and provides an effective particle size. The Voronoi cell volume is $V_{c(ell)}$.
For monodisperse packings, $A=1$. For the bidisperse packings, $A$ is different for the two particle types. The average squared effective particle size is $\langle D_v^2\rangle=m (D_{V,big})^2+(1-m)(D_{v,small})^2$ where $m=2/5$ is the ratio of big particles. $A_{big}=(D_{v,big})^2/\langle D_v^2\rangle$ and $A_{small}=(D_{v,small})^2/\langle D_v^2\rangle$. These are the same as in the paper by Yuan {\it et al.}, $v_f=v_f(s_1,s_2)$ and $A=A(s_1)$. 

We find values that our data $v_f(A)$ are consistent with but slightly higher than Yuan {\it et al}. This is consistent with our particles being frictional, albeit only slightly, which increases $v_f$ while not affecting $A$.  Evidently, as compactification through tapping affects $v_f$ but not $A$, our curves for different values of $\langle \phi_l\rangle$ remain distinct, with the most compacted packings approaching the curve for the frictionless particles studied by Yuan {\it et al}. The data is presented in Fig.\ \ref{fig:Yuan-averages}.
}

\section*{Author Contributions}

Authors contributions to this work are as follows: Conceptualisation: FMS, MS, GEST; Experiments: FMS, HP; Data curation: FMS; Formal analysis: FMS, GEST, MS; Funding acquisition: GEST; Investigation: all authors; Methodology: GEST, FMS and MS; Project administration: GEST and MS; Resources: FMS, HP, MS; Software: FMS; Supervision: GEST and MS; Validation: FMS and GEST; Visualisation: FMS and GEST; Writing: FMS, GEST and MS; Reviewing: all authors.

\section*{Conflicts of interest}
There are no conflicts to declare.

\section*{Acknowledgements}
We thank Tim Senden and Sebastian Kapfer for his insightful comments and fruitful discussions. We thank Weimer Pharma GmbH for the provision of the placebo pills. We acknowledge funding by the German Science Foundation (DFG) through the research group ``Geometry and Physics of Spatial Random Systems''. FMS acknowledges funding by the Marcelja Fellowship of the Department of Applied Maths at the Australian National University. MS acknowledges financial support from the Australian Research Council (ARC) under project IC180100008. GEST acknowledges current funding by the Australian Research Council (ARC) through the Discovery Project DP200102593.



\balance


\providecommand*{\mcitethebibliography}{\thebibliography}
\csname @ifundefined\endcsname{endmcitethebibliography}
{\let\endmcitethebibliography\endthebibliography}{}

\end{document}